\documentclass[prb,aps,twocolumn,floatfix,amsmath,amssymb,superscriptaddress,tightenlines]{revtex4}
\usepackage{graphicx}
\usepackage{epstopdf}
\usepackage{amsfonts}
\usepackage{bm}
\usepackage{color}
\usepackage{ulem}
\begin{document}

\newcommand{\be}{\begin{equation}}
\newcommand{\ee}{\end{equation}}

\date{\today}
\title{Entanglement scaling in two-dimensional gapless systems}

\author{Hyejin Ju}
\affiliation{Department of Physics, University of California, Santa Barbara, Santa Barbara, California, 93106-9530, USA}

\author{Ann B. Kallin}
\affiliation{Department of Physics and Astronomy, University of Waterloo, Waterloo, Ontario N2L 3G1, Canada} 

\author{Paul Fendley}
\affiliation{Physics Department, University of Virginia, Charlottesville, Virginia 22904-4714, USA}
\affiliation{Microsoft Research, Station Q, CNSI Building, University of California, Santa Barbara, Santa Barbara, California, 93106-9530, USA}

\author{Matthew B. Hastings}
\affiliation{Microsoft Research, Station Q, CNSI Building, University of California, Santa Barbara, Santa Barbara, California, 93106-9530, USA}

\author{Roger G. Melko}
\affiliation{Department of Physics and Astronomy, University of Waterloo, Waterloo, Ontario N2L 3G1, Canada} 
\affiliation{Perimeter Institute for Theoretical Physics, Waterloo, Ontario N2L 2Y5, Canada}

\begin{abstract} 
We numerically determine subleading scaling terms in the ground-state entanglement entropy of several two-dimensional (2D) gapless systems, including a Heisenberg model with N\'eel order, a free Dirac fermion in the $\pi$-flux phase, and the nearest-neighbor resonating-valence-bond wavefunction.
For these models, we show that the entanglement entropy between cylindrical
regions of length $x$ and $L-x$, extending around a torus of
length $L$, depends upon the dimensionless ratio $x/L$.  This can be well-approximated on finite-size lattices by a function
$\ln(\sin(\pi x/L))$, akin to the familiar chord-length dependence in one dimension. 
We provide evidence, however, that the precise form of this bulk-dependent contribution
is a more general function in the 2D thermodynamic limit.

\end{abstract}
\maketitle

{\it Introduction --} 
The study of quantum condensed matter systems
 is benefiting from an infusion of ideas related to quantum information and entanglement. The importance of this new resource is strikingly 
demonstrated in the study of entanglement entropy at one-dimensional (1D) quantum critical points with conformal invariance. Conformal field theory (CFT) provides an important
universal number, the {\it central charge} $c$, which appears in an astonishing array of physical
quantities.\cite{Cardyubiquitous} A given CFT, and
thus any quantum critical points it describes, can be
characterized by this number.
Its numerical or analytical determination provides an invaluable tool for identifying which, if any, CFT describes the scaling limit of a given Hamiltonian. 
Computing the entanglement entropy has proven to be a very useful way of finding $c$ numerically. It can be extracted 
directly from the ground-state wavefunction by measuring its Renyi
entanglement entropy, $ S_n = 1/(1-n) \ln \big[ {\rm Tr} \rho_A^n
\big], $ where region $A$ is entangled with its complement, region
$B$. Namely, in a system with total length $L$, where the region $A$ has length $x$, the scaling of the Renyi entropy in 1D critical systems depends on the ``chord length" as,\cite{Holzhey,VidalC,Korepin,Cardy}
\begin{equation}
S_n = \frac{c}{6}\left({1+ \frac{1}{n} }\right) \ln\Big[ \frac{L}{\pi} \sin\big( \frac{\pi x}{L} \big) \Big], \label{1Dcft}
\end{equation}
with the central charge appearing as the coefficient.

In higher dimensions, the scaling behavior of the entanglement entropy
is much less well understood.  Ground states of local Hamiltonians are
generally believed to produce an ``area-law'' (i.e.\ boundary) scaling,
\cite{ALreview} the subleading corrections to which may be universal
quantities that can be used to identify and characterize quantum
phases and phase transitions.  A well-established example of this is the 
topological entanglement entropy\cite{Alioscia1,Alioscia2,KP,LW} of
a gapped state with topological order.  In gapless states, the
subleading corrections may still potentially harbor universal
quantities. It is conceivable that such quantities could be used to define an
``effective'' central charge in two spatial dimensions, but there are strong constraints on any proposal.\cite{EE_CFT} The best-understood gapless situation in two dimensions is the special case of a conformal quantum critical point,
where the ground state itself is written in terms of a two-dimensional (2D)
CFT.\cite{Moore06,Hsu08,Misguich,Oshikawa,Hsu10,Zaletel,Stephan11} In the presence of a spontaneously broken
continuous symmetry, Goldstone modes produce a subleading bulk
logarithmic correction.\cite{HeisLog,MaxLog}  Subleading logarithms
from corner contributions with universal coefficients also occur at
some critical points.\cite{Moore06,logcorner,Max}  

The purpose of this
paper is to analyze one type of subleading term in 2D gapless systems
and to study whether this term is universal. Gapless modes typically have long-range correlations, so it is possible for
the entanglement entropy to depend on the size and shape of regions $A$ and $B$. Indeed, the 1D result,
(\ref{1Dcft}), is manifestly size dependent. We show how similar behavior also occurs in two deimensions.

\begin{figure}
   \begin{center}
   \scalebox{1}{\includegraphics[width=2.1in]{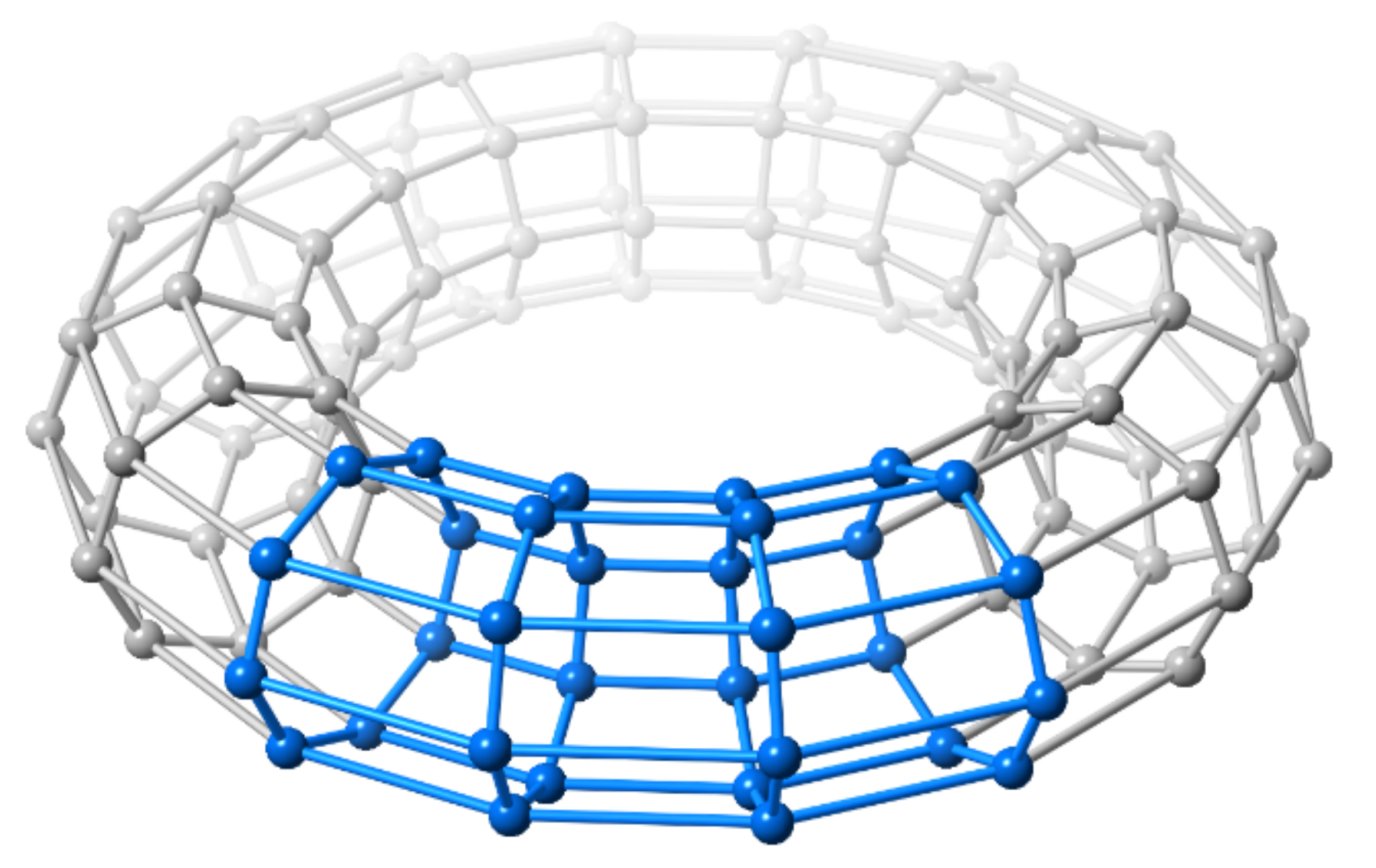}}
   \end{center}
   \caption{(Color online) An $8 \times 16$ toroidal lattice.  The width of cylindrical region $A$ (blue) is $x=4$.  The boundary length between region $A$ and its complement is $\ell = 16$. }
   \label{fig:torus}
 \end{figure}

We study the finite-size scaling of the second Renyi entropy for
the ground states of several 2D gapless systems on the square lattice using quantum Monte Carlo (QMC) simulations. It is possible to vary the size of regions $A$ and $B$ without changing the length of the boundary between a toroidal lattice geometry, where $A$ and $B$ are cylinders as in Fig.\ \ref{fig:torus}. 
We examine the N\'eel ground state of the Heisenberg model, and the nearest-neighbor resonating-valence-bond (RVB) wavefunction, in this geometry.
In both cases, we find a size- and shape-dependent scaling function that closely mimics the chord-length contribution in one dimension in Eq.~(\ref{1Dcft}).

To probe this behavior in a simpler system, we also study
 free spinless fermions in the $\pi$-flux phase and find that the
 entanglement scaling also has a universal size- and shape-dependent piece.
For finite-size systems, this closely mimics the chord length, but in the infinite-size limit we observe it to cross over to a different nontrivial function.
Among other consequences, this term will give a nonzero signature in the entanglement quantities\cite{KP,LW} designed to look for topological order, which complicates any possible generalization of the topological entanglement entropy to gapless spin-liquid states. 
 
{\it Fermions with $\pi$ flux---}
We begin by considering
free spinless fermions on a square lattice, with $\pi$ flux through each plaquette.  We consider a torus of size $L_x$ by $L_y$, and
measure the entanglement  
using a cornerless cylindrical region $A$ (Fig.~\ref{fig:torus}) with a constant boundary length $\ell = 2L_y$.
We denote the width of region $A$ by $x$.
This system has Dirac points near momentum $k_y=0$ and $k_y=\pi$.  We take antiperiodic boundary conditions in the $x$ direction so that there will be no exact zero mode.  We use exact numerical diagonalization of the single-particle Hamiltonian to compute the entropy.
The entanglement entropy of 2+1-dimensional conformally invariant systems such as this has been argued to be of the form,\cite{ryu,ZGV} 

\begin{equation}
S_n \sim {\rm const.}\times \ell /a + \gamma(x/L_x,L_y/L_x),
\end{equation}
where $\gamma$ is a universal scaling function of the dimensionless ratios.
The area-law term proportional to the boundary length $\ell$ depends on the lattice constant $a$, and so the constant is nonuniversal. A crucial difference
from the result in one dimension, Eq.~(\ref{1Dcft}), is that $a$ only appears in the area-law term.  In contrast, the 1D result can be written as the sum of two terms, as
$S_n =C \ln[\sin\big( \frac{\pi x}{L} \big)]+ C\ln[\frac{L}{\pi}] $, where $C=c/6(1+1/n)$.
The first term is a universal function of the dimensionless ratio $x/L$, akin to the function $\gamma$ above, while the second term involves the lattice scale, as it diverges with $L$.

To illustrate the absence of such an ``additive logarithm" (a logarithmic divergence depending on $L/a$) in two dimensions,
we treat this free system as a collection of independent systems in one dimension labeled by the momenta $k_y$.
The $k_y=0$ mode contributes an additive logarithm $C \ln(L_x)$ to the entropy, while the modes with small $k_y \neq 0$ contribute additive logarithms $C \ln(k_y^{-1})$.\cite{Holzhey,Korepin,Cardy} Summing over $k_y=2\pi j/L_y$, this gives an entropy $C \big[ \ln(L_x)+2\sum_{j=1}^{j \sim L_y} \ln(L_y/2 \pi j) \big]=C\ln(L_x) + 2 C\ln[(L_y/2\pi)^{L_y}/L_y!]$, where the factor of $2$ arises from summing over positive and negative $m\neq 0$.  Using Stirling's formula for $L_y!$, one finds that the additive logarithm terms add to $C[\ln(L_x)-\ln(L_y)]=C\ln(L_x/L_y)$. This can be absorbed into the scaling function $\gamma$, so that there is no additive logarithm.  A more precise calculation would include the effect of finite $L_x$, but we ignore this since it does not affect the cancellation of additive logarithms. A similar calculation near $k_y=\pi$ leads to a cancellation of the additive logarithm there.

\begin{figure}
   \begin{center}
   \scalebox{1.7}{\includegraphics[width=2.1in]{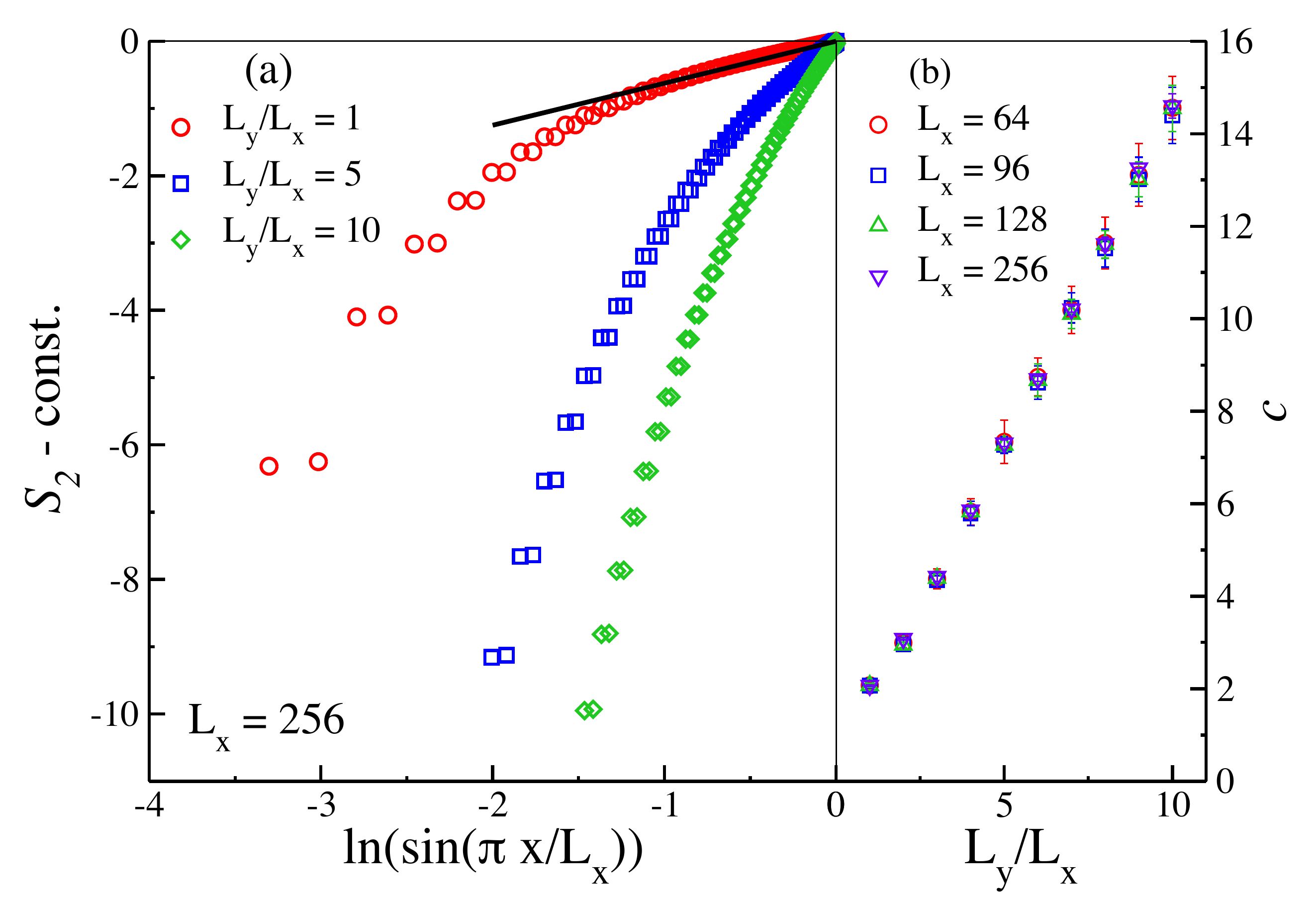}}
   \end{center}
   \caption{(Color online) (a) Renyi entropy of the $L_x=256$ Dirac fermion model, with an arbitrary constant subtracted off each data set.
    The straight line is Eq.~(\ref{1Dcft}) plotted with $c=2$.
    It is clear that deviations from linearity increase for larger $L_y/L_x$. (b) A set of points near $x = 0$ is shown to follow the 1D result $\ln(\sin(\pi x/L_x))$. Note that the ``central charge," $c$, in this region is independent of the system size.
   }
   \label{fig:dirac}
 \end{figure}

The entropy of a given $k_y$ mode contains, in addition to the additive logarithmic divergence in $k_y$, a universal scaling function $G(x/L_x,k_y x)$.  At $k_y=0$, we see the chord-length scaling $C \ln \big[\sin ( \frac{\pi x}{L} ) \big]$ (Fig.~\ref{fig:dirac}), but for $k_y \neq 0$ and for $k_y x$ large, the chord-length scaling disappears and the entropy becomes roughly flat as a function of $x/L$.  In fact, for $L_y=L_x=L$, the lowest $k_y$ mode has a mass $2 k_y=4\pi/L$. This factor of $4\pi\approx 13$ means that this mass is rather large, and so the entropy of this mode is flat for a large range of $x/L$.  As a result, for $L_y=L_x=L$, the entropy of the 2D system appears to display 1D chord length scaling over a wide range of $x/L_x$.

\smallskip 
{\it Quantum Monte Carlo---}
Using QMC techniques, we simulate both the
Heisenberg ground state and the RVB
wave function in two dimensions.  The Heisenberg ground state is
projected from a trial state by applying a high power of the
Hamiltonian, $H = \sum_{\langle ij \rangle} {\bf S}_i \cdot {\bf S}_j$, via a QMC method operating in the valence bond (VB)
basis.\cite{Sandvik}
The RVB wavefunction 
is an equal-amplitude superposition 
$| \Psi \rangle = \sum_{\alpha} | V_{\alpha} \rangle$
of all nearest-neighbor VB states, 
\begin{equation}
|V_{\alpha} \rangle =  \frac{1}{2^{N/4}} \prod_{i=1}^{N/2} \big( | \uparrow_i \downarrow_{j_{\alpha}} \rangle -  | \downarrow_i  \uparrow_{j_{\alpha}} \rangle \big),
\end{equation}
 defined by requiring that each spin $i$ on one sublattice be in a singlet with one of its nearest neighbors $j_\alpha$.\cite{RVB1,RVB2} The RVB 
Monte Carlo sampling
algorithm does a random walk through the possible states by creating a
defect at some spatial point and propagating it through the system (thereby
rearranging the nearest-neighbor bonds) until the defect reaches the
initial point and its path forms a closed loop.\cite{AWSloop}
If we visualize the Heisenberg ground state in this VB language, then the RVB wave function is its largest component, the remainder of the state being superpositions of longer bonds, decaying with their length as $1/r^3$.\cite{Sandvik}  Likewise, the RVB wave function is the ground state of a local (but longer range) Hamiltonian that includes a Heisenberg term.\cite{Cano}

We consider the same geometry as for the $\pi$ flux fermions, with $L_x=L_y=L$ (see Fig.~\ref{fig:torus}).
 \begin{figure}
   \begin{center}
   \scalebox{1}{\includegraphics[width=0.9\columnwidth]{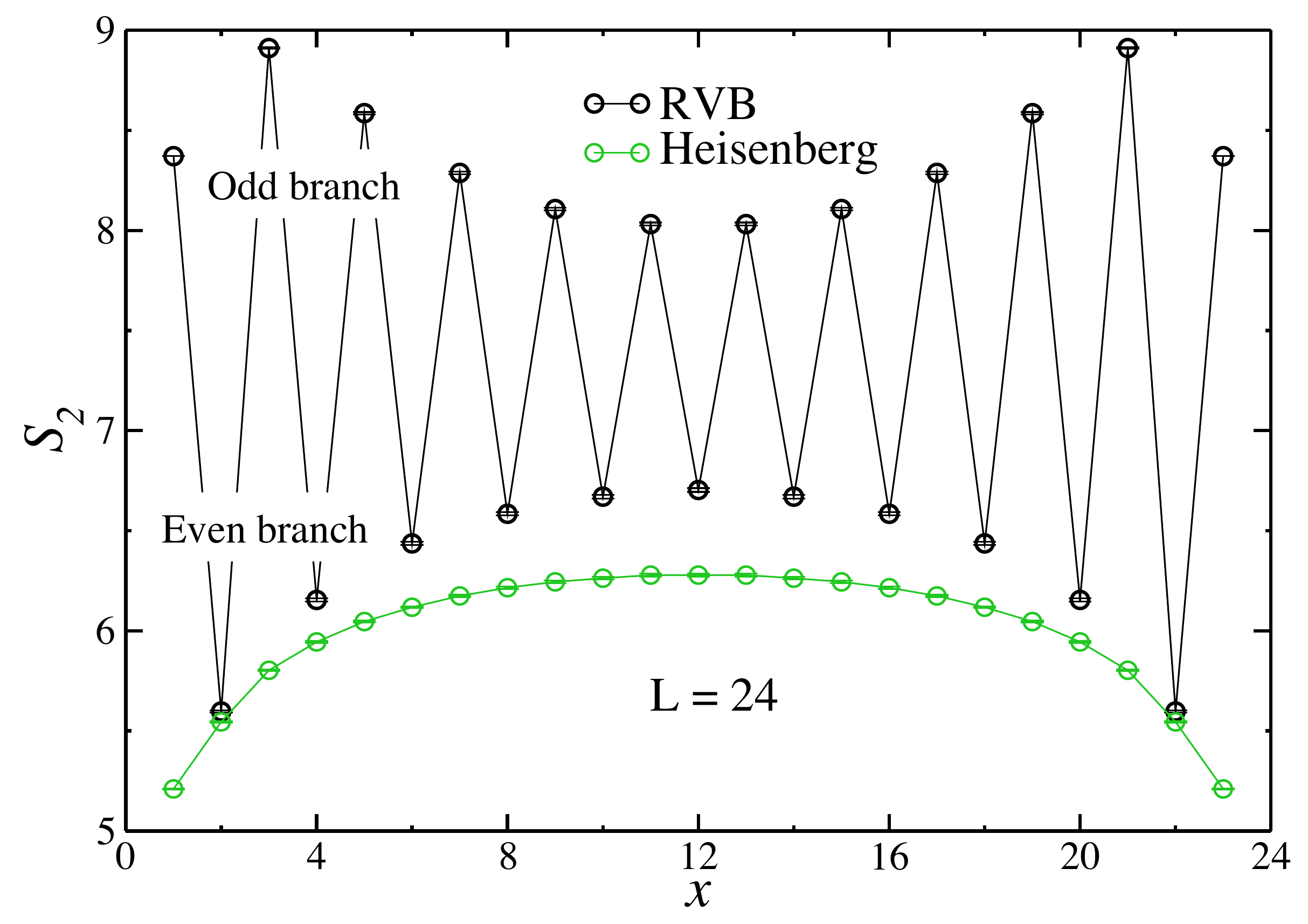}}
   \end{center}
   \caption{(Color online) The second Renyi entropy for the N\'eel and RVB states for $L=24$. Note that the entanglement entropy for the RVB splits into two branches, even and odd, which may be related to the existence of topological sectors in the underlying transition graphs.
   \label{fig:heis_bow}}
 \end{figure}
In Fig.~{\ref{fig:heis_bow}}, we plot QMC results for the second Renyi entropy in the N\'eel and RVB states on a $24 \times 24$ torus.  
Several features of the entanglement scaling are clear from this plot.  First, note that the data for the N\'eel state has a significant curvature as a function of $x$.  This curvature was first seen in Ref.~[\onlinecite{HeisLog}] but not explored in detail [instead, using a fixed $x/L$, a surprising subleading logarithmic term $\propto \ln(\ell)$ was found\cite{MaxLog}].  
The entropy of the RVB wave function exhibits an obvious dependence on whether $x$ is even or odd which we discuss in more detail below. In each of the even and odd ``branches," 
there is significant curvature in the $x$ dependence as with the Heisenberg case. 

\begin{figure}
   \begin{center}
   \scalebox{1}{\includegraphics[width=0.95\columnwidth]{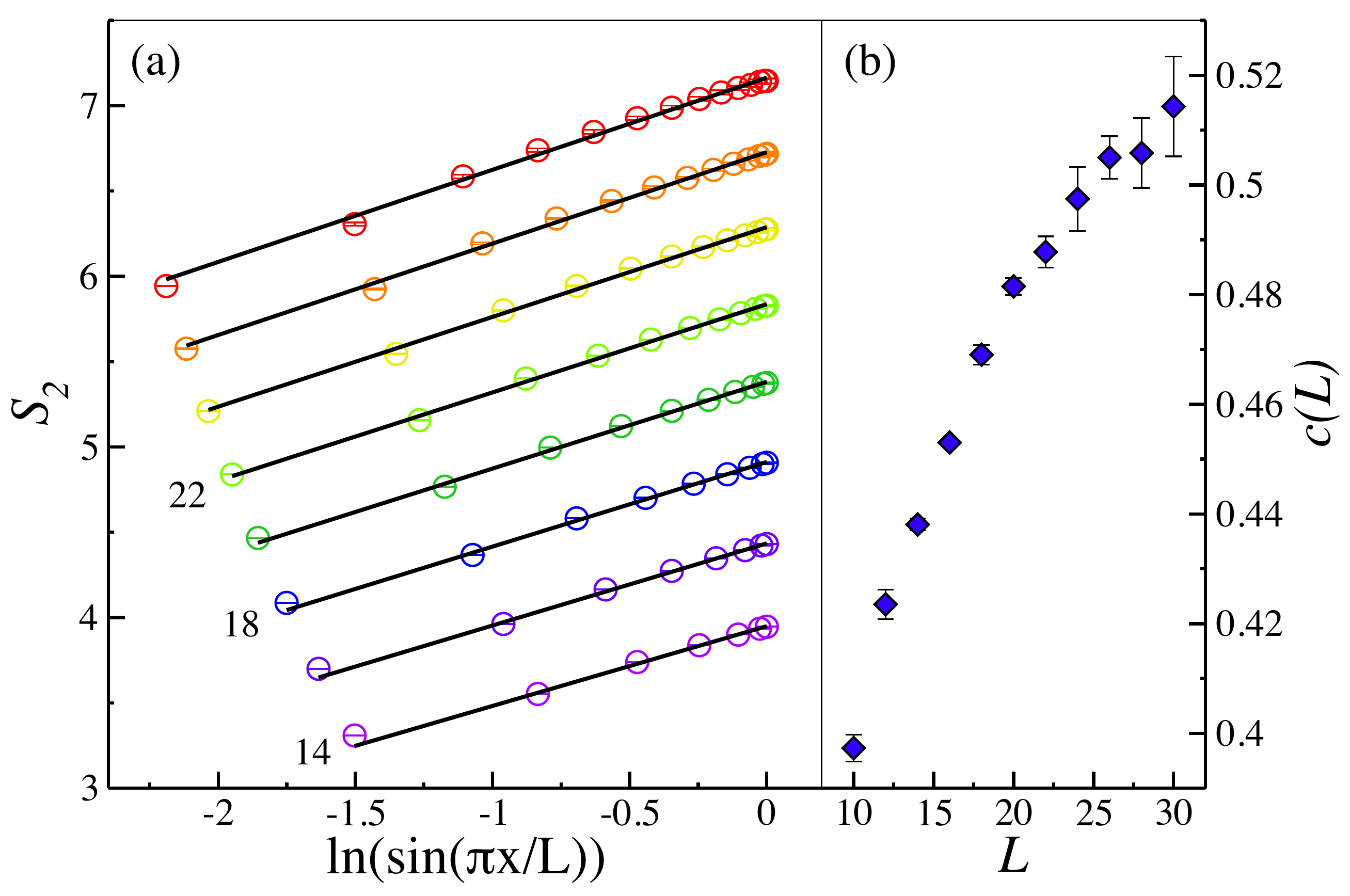}}
   \end{center}
   \caption{(Color online) (a) Heisenberg data and linear fits (excluding the first two data points on the left) for $L=14,16,\dots,28$ plotted in terms of the log of the ``chord length," $\ln\left[\sin \frac{\pi x}{L}\right]$.
   (b) Slopes of the fits, $c(L)$, exhibit a strong dependence on the system size, $L$.
   }
   \label{fig:heis_lines}
 \end{figure}

To capture the $x$-dependent curvature of these wave functions, we fit the data with the scaling ansatz,
\begin{align}
S_2= a \ell + b\ln(\ell)
+ c(L) \ln \left[{ \sin\left({ \frac{\pi x}{L} }\right) }\right] + d, \label{Fit}
\end{align}
motivated by the chord length in Eq.~(\ref{1Dcft}).
We begin by examining the N\'eel state in Fig.~{\ref{fig:heis_lines}}.  For a fixed linear system size $L$ and boundary length $\ell$, plots of $S_2$ versus $ \ln \left[{ \sin\left({ \frac{\pi x}{L} }\right) }\right] $ would yield a straight line if Eq.~(\ref{Fit}) were obeyed perfectly.  The plots indeed are quite close to straight lines for a fixed $L$.

The second Renyi entropy therefore displays, at the very least, an effective chord-length dependence over a large range of $x$ for the square torus. It is possible that the apparent chord-length scaling of this 2D system is not perfectly obeyed in the thermodynamic limit and that this fact is manifest in slight deviations from straight-line behavior in Fig.~{\ref{fig:heis_lines}}(a).
This would be a similar scenario to the deviation from chord-length scaling observed for $\pi$-flux fermions in Fig.~\ref{fig:dirac}.  
However, it is difficult to draw a firm conclusion regarding the statistical significance of any deviation from Eq.~(\ref{Fit}) scaling in our
present data, due to limited system sizes and stochastic error.

We can, however, further examine the deviation from conformal-style scaling by extracting the $L$ dependence of the coefficient $c(L)$ in Eq.~(\ref{Fit}).  In order for this shape-dependent term to be universal in two dimensions, $c(L)$ should approach a constant in the limit $L \rightarrow \infty$ for fixed $x/L$.
As illustrated in Fig.~\ref{fig:heis_lines}(b),
the coefficient does not approach a constant for the system sizes that we have studied but, rather, has some functional dependence on $L$.
This functional dependence is apparently sub linear - possibly behaving like $c(L) \sim L^p$ with $p\leq1$.  This scenario could be supported by the QMC data if convergence were assumed to be very slow. Indeed, in the quantum dimer model, the corresponding term can be computed exactly in finite size, and the convergence is very slow.\cite{Stephan12} A definitive determination of this limit (and therefore the strict adherence of $\gamma$ to universality in this system) is impossible with our current data; significantly larger system sizes must be studied.

 \begin{figure}
   \begin{center}
   \scalebox{1}{\includegraphics[width=\columnwidth]{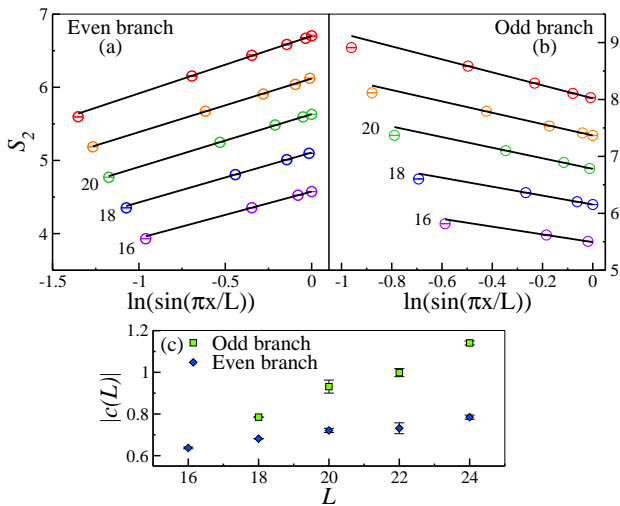}}
   \end{center}
   \caption{(Color online) (a) Even and (b) odd branches of the second Renyi entropy plotted against the log of the ``chord length." We exclude $x=1,2$ data from the plots, as there is some crossover behavior shown in Fig.~\ref{fig:heis_bow}. (c) Absolute value of the slopes, $|c(L)|$, as a function of the system size, $L$. As with the Heisenberg, there is a strong dependence on $L$.}
   \label{fig:2}
 \end{figure}

We next examine the scaling of the Renyi entropy in the RVB wave function.
As shown in Fig.~\ref{fig:heis_bow}, a striking two-branch structure exists, depending on whether the distance $x$ is even or odd.
The presence of the two branches presumably is related to the fact that correlators in the RVB state have a pronounced even-odd dependence. Moreover, simple counting arguments of prototypical VB configurations in the (0,0) topological sector\cite{RVB1,RVB2} show that the number of VBs crossing from region $A$ to region $B$ alternates strongly with $x$.  This $L=24$ data display a clear $x$-dependent curvature in each branch. This can be analyzed more closely by attempting fits of the form of Eq.~(\ref{Fit}) to each branch individually. From Figs.~{\ref{fig:2}}(a) and~{\ref{fig:2}}(b), it is clear that fits to the scaling
ansatz to both branches are quite accurate when the extremal values of
$x$ are excluded. 
It is worth noting that in the closely related quantum dimer
model on a square lattice, similar terms appear and can be computed exactly,\cite{Stephan12} generalizing the results of Refs.~[\onlinecite{Hsu08,Misguich,Oshikawa,Hsu10,Zaletel,Stephan11}].

We can attempt to
extract the size dependence of the coefficient $c$ in a similar
manner as for the Heisenberg model. Unfortunately, due to the two-branch structure, each curve in this plot has
essentially half the usable data compared to the analogous Heisenberg
results in Fig.~\ref{fig:heis_lines}.  Nonetheless,  the result [Fig.~\ref{fig:2}(c)] shows that a significant
$L$ dependence seems to exist in the RVB wave function as well.  This
again suggests that, although the fit to a chord-length scaling at
fixed $L$ is consistent within the accuracy of our data, subtle
corrections to this form may come into play in the 2D thermodynamic
limit. 
\smallskip

{\it  Discussion---} We have studied the Renyi entanglement entropy in the ground state of three gapless systems on $L_x\times L_y$ toroidal lattices, where the subregion $A$ is a cylinder of length $x$.  
We have demonstrated that it contains a subleading scaling term which depends on bulk quantities, namely the dimensionless aspect ratios of the subregion and the lattice linear dimensions.
Note that while numerical measurement of topological entanglement entropy\cite{LW,KP} has been used to probe topological properties of {\it gapped} phases,\cite{isakov} the subleading term considered here means that a measurement of topological entanglement entropy in a {\it gapless} phase could give either a zero or a nonzero result, even without any topological aspects of the phase [though measurements in the $U(1)$ superfluid phase yielded a vanishing number\cite{isakov}].  
Interestingly, just as strong subadditivity constrains the sign of the Levin-Wen entropy,\cite{LW} it also implies, for fixed $L_x,L_y$, that $\gamma$ for the von Neumann entropy is a concave-down function of $x$.

Our QMC simulations of the Heisenberg N\'eel ground state and the short-range RVB wave function with $L_x=L_y=L$ show an almost-perfect logarithmic dependence of $\gamma$ on the chord length $\sin(\pi x/L)$.
It appears that the coefficient of this term is not a universal constant, however, which might suggest either that care must be taken in the order of limits with which the thermodynamic limit is approached or that a size dependence remains in this limit, rendering this term nonuniversal. A study of the crossover from one to two dimensions might illuminate this issue further.  Further evidence that the true 2D scaling function might not be exactly the chord-length form is given by the scaling of gapless Dirac fermions in the $\pi$-flux phase. Here we have argued that such scaling is superseded by a sum over transverse modes, leading to a different (unknown) functional form in two dimensions.  Furthermore,  spontaneous symmetry breaking in the Heisenberg model may complicate measurement of the entanglement entropy.  The fact that a complete characterization of the scaling behavior in the N\'eel and RVB states remains a challenge, despite the large lattice sizes studied to date, underlines the absolute necessity for using large-scale QMC simulations for the study of entanglement entropy.

Regardless of the precise functional form of 
the shape-dependent subleading term $\gamma$,
its general existence in gapless wavefunctions in two dimensions would have some profound consequences.  
Besides the immediate complications in attempting to use entanglement as a probe to detect gapless spin liquids mentioned above,
the similarity of the scaling function to a chord length (present in 1D conformally invariant systems) 
raises the tantalizing possibility that our results will prove useful in characterizing higher dimensional critical points.
Indeed, since the search for a $c$ theorem\cite{Zamo} valid in higher dimensions is of intense interest across several disparate field of physics,\cite{Cardy88,ryu,Myers,Komargodski}
we hope that our results will inspire a broader examination of this scaling term in 2D gapless states.

{\it Acknowledgments---} 
The authors thank J.-M. St\'ephan, A.~Del Maestro, E.~Fradkin, I.~Klich, K.~Intriligator,  M.~Metlitski, E. G.~Moon, R.~Myers, and A.~Sandvik for enlightening discussions. 
R.G.M. would like to acknowledge the support of Microsoft Station Q for hospitality during a visit.
This work was supported by the Natural Sciences and Engineering
Research Council of Canada, and by the U.S. National Science Foundation via Grants Nos. DMR/MPS-0704666 and DMR/MPS1006549.  Simulations were performed on the computing facilities of SHARCNET.

\end{document}